\documentclass[twocolumn]{aastex62}

\submitjournal{ApJL}
\accepted{June 22, 2019}

\usepackage{hyperref}

\usepackage[switch]{lineno}
\usepackage{color}
\usepackage{ulem}

\begin{document}

\title {Record-breaking coronal magnetic field in solar active region 12673 }

\correspondingauthor{Sergey A. Anfinogentov}
\email{anfinogentov@iszf.irk.ru}

\author[0000-0002-1107-7420]{Sergey A. Anfinogentov}
\affil{Institute of solar-terrestrial physics, Lermontov st., 126a, Irkutsk 664033, Russia}

\author[0000-0002-5453-2307]{Alexey G. Stupishin}
\affiliation{Saint Petersburg State University, 7/9 Universitetskaya nab., St. Petersburg, 199034 Russia}

\author[0000-0002-8530-7030]{Ivan I. Mysh'yakov}
\affiliation{Institute of solar-terrestrial physics, Lermontov st., 126a, Irkutsk 664033, Russia}

\author[0000-0001-5557-2100]{Gregory D. Fleishman}
\affiliation{Center for Solar Terrestrial Research, New Jersey Institute of Technology, University Heights, Newark, NJ 07102-1982, USA}

\begin{abstract}
At the Sun, the strongest magnetic fields are routinely detected at dark sunspots. The magnitude of the field is typically about 3000\,G, with only a few exceptions that reported the magnetic field in excess of 5000\,G. Given that the magnetic field decreases with height in the solar atmosphere, no coronal magnetic field above $\sim$2000~G was ever reported. Here, we  present imaging microwave observations of anomalously strong  magnetic field of about 4000~G at the base of the corona in solar active region NOAA 12673 on 06 September 2017.
Combining the photospheric vector measurements of the magnetic field and the coronal probing, we created and validated a nonlinear force-free field coronal model, with which we quantify the record-breaking coronal magnetic field at various coronal heights. 
\end{abstract}

\section{Introduction}
At the Sun, the strongest large-scale magnetic field is concentrated at certain areas called active regions (ARs). The magnitude of the field is typically about 3000\,G, with only a few exceptions that reported the magnetic field in excess of 5000\,G \citep{1993SoPh..144...37Z, 2006SoPh..239...41L, 2016ApJ...818...81J, 2018ApJ...852L..16O, 2018RNAAS...2a...8W}. In their rigorous survey of historical records covering almost 90 years of measuring photospheric magnetic field in ARs, \citet{2006SoPh..239...41L} report 55 ARs with magnetic field above 4,000\,G, 5 cases above 5,000\,G, and even one case above 6,000\,G. In particular, AR~7378 had a magnenic field of $\sim$6,100\,G on 1942 Feb 28. This AR was associated with a strong geomagnetic storm and gave rise to discovery of sporadic solar radio emission in the meter-wave band \citep{2006SoPh..239...41L}. A  strong magnetic field is needed to drive extreme solar events; however, given the rarity of the strongest field cases, no reliable association rate has yet been established.

The ARs are called this way because they control most of the solar eruptive activity such as solar flares or coronal mass ejections. This activity is powered by the coronal magnetic field, whose energy dominates other forms of energy in the corona. Thus, it is fundamentally important to carefully quantify the coronal magnetic field in ARs. However, 
no routine diagnostics of the coronal magnetic field at different heights are currently available, so various modeling approaches are used instead. 

An exception is the gyroresonant (GR) probing of the coronal magnetic field, which is performed with microwave imaging instruments. The foundation of the method is very simple. 
The coronal plasma is optically thick at a few lowest harmonics of the local gyro-frequency. The brightness temperature of this optically thick emission is equal to the kinetic temperature of the emitting plasma. The stronger the local magnetic field the higher the radio frequency at which the given source is bright. 
For a given line of sight,  the magnetic field increases towards the solar surface, while the temperature decreases sharply at the transition region between the corona and the chromosphere. Thus, the radio brightness will drop sharply at the emission frequency that corresponds to the gyroresonant frequency at the transition region. Investigating the radio brightness pixel-by-pixel, we can recover the magnetic field strength at the base of the corona, provided that broad-band microwave data with high spatial resolution are available. 

In practice, this diagnostics is often available either at a few single frequencies 
or over a limited spectral range.                                                                  
For a typical AR, the GR emission dominates below $\sim$10~GHz, while stronger ARs show bright GR emission at 17--18~GHz \citep{2007SSRv..133...73L, 2011SoPh..273..309S}. No GR emission at higher frequencies was ever reported implying that the coronal magnetic field does not typically exceed $\sim$2100~G. 

However, AR 12673 demonstrated an extremely strong photospheric field of about 5,000\,G over a rather extended area of the sunspot during the day of 06 September 2017.
On top of that, a light bridge at the polarity inversion line demonstrated at times extreme values of the transverse magnetic filed up to 5,570\,G \citep{2018RNAAS...2a...8W}.
Magnetic field of such magnitude was formed 
in association with an unusually fast
new magnetic flux emergence \citep{2017RNAAS...1a..24S}. During the period of 03 September to 04 September, a number of magnetic bipoles with a scattered structure subsequently emerged on the eastern side of the pre-existing sunspot. Their complex mutual motion leads to the formation of the high sheared strong magnetic configuration that eventually produced an X9.3 flare on September 06. A comprehensive description of the photospheric magnetic field evolution of AR 12673 can be found in \citet{2017ApJ...849L..21Y}. \textbf{[Some text was removed here.]}
Although this strongest transverse magnetic field may close very low at the chromosphere and have no appreciable fingerprint in the corona, the more extended magnetic field of the order of 5,000\,G may reach the higher coronal heights.
If this is the case, a bright gyroresonant (GR) coronal source might appear at an unusually high microwave frequency.

Here, we report on such a bright GR source at the highest available frequency, 34\,GHz, using Nobeyama RadioHeliograph (NoRH) data during the day of 06-Sep-2017, which is  twice larger than the largest frequency at which a GR source has ever been reported. This observation {\bf alone} unambiguously indicates the presence of the coronal magnetic field of  up to 4000\,G which is twice larger than values reported so far. 

\section{Observations and Data Analysis}

\subsection{Overview.} 
In this study we employ spectropolarimetry data obtained with the Helioseismic and Magnetic Imager  Instrument on the Solar Dynamics Observatory  \citep[SDO/HMI,][]{2012SoPh..275..229S,2012SoPh..275....3P} and Hinode SOT/SP \citep{2008SoPh..249..167T,2007SoPh..243....3K} to derive the photospheric vector magnetograms needed to initiate coronal magnetic models as well as microwave images obtained with Nobeyama RadioHeliograph \citep[NoRH,][]{1994IEEEP..82..705N} to quantify the coronal magnetic field.

Given that AR 12673 demonstrated remarkably unusual behavior, many standard pipeline products contain artifacts, while  standard analysis techniques  required manual corrections as described below. For our analysis, we focused on two time frames: 03:34:42~UT and  at 18:36:50 UT on 6 September 2017. The first time frame is selected during local mid-day of the microwave observations at the Nobeyama Observatory that detected a unique long-living gyroresonant source at 34\,GHz, while the amount of SDO/HMI artifacts was minimal during that time period. The second time frame is selected based on the time when the largest photospheric magnetic field was detected by Hinode SOT/SP.

\subsection{Inversion of the Hinode SOT/SP Stokes profiles.}

NOAA AR 12673 has been observed by both SDO/HMI and Hinode SOT/SP instruments.
In the standard vector magnetograms available at \url{http://jsoc.stanford.edu/} for SDO/HMI and at \url{https://csac.hao.ucar.edu/sp_data.php} for Hinode SOT/SP, the magnetic field is artificially bounded below 5000\,G. For this active region, the magnetic field is saturated at this threshold. 

\begin{figure*}[!t]
\centerline{\includegraphics[width=0.9\linewidth]{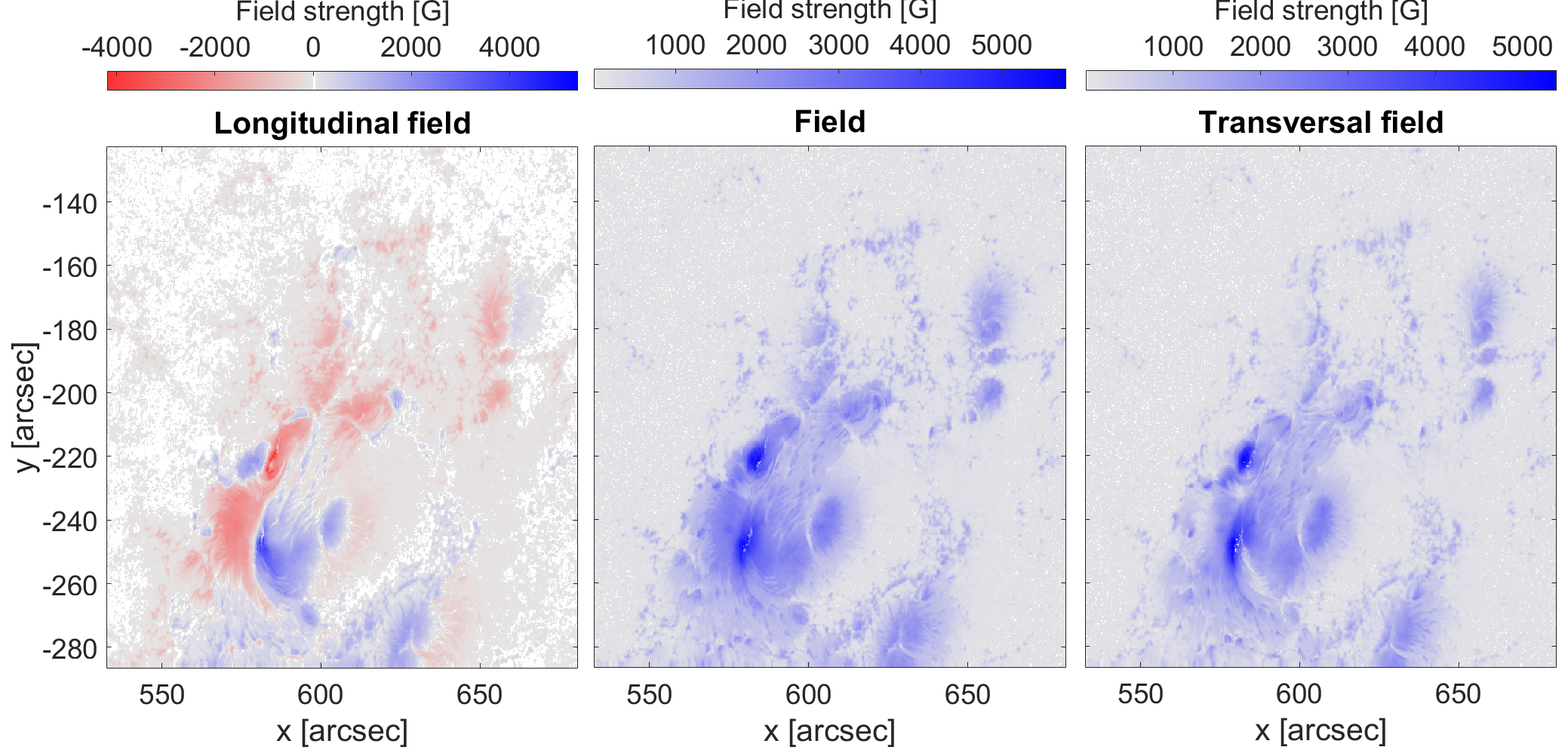}}
	\caption{
	Map of the absolute value of the magnetic field (central panel), its longitudinal (left panel) and transversal (right panel) components observed by Hinode SOT/SP at 18:36:50 UT on 6 September 2017. }
    \label{fig:hinode_1836}
\end{figure*}

To measure the real (unsaturated) value of the magnetic field from Stokes profiles observed by Hinode SOT/SP at 18:36:50 UT on 6 September 2017, we use the VFISV inversion code \citep{2011SoPh..273..267B} available at \url{https://www2.hao.ucar.edu/csac/csac-spectral-line-inversions}.
We modified this source code to allow magnetic fields above 5000\,G and applied to the measured Stokes profiles.
This way we obtained unsaturated vector magnetogram. The corresponding maps of the absolute value, as well as longitudinal and transversal components of the magnetic field vector are shown in  Figure \ref{fig:hinode_1836}.

\begin{figure}
\includegraphics[width=0.95\linewidth]{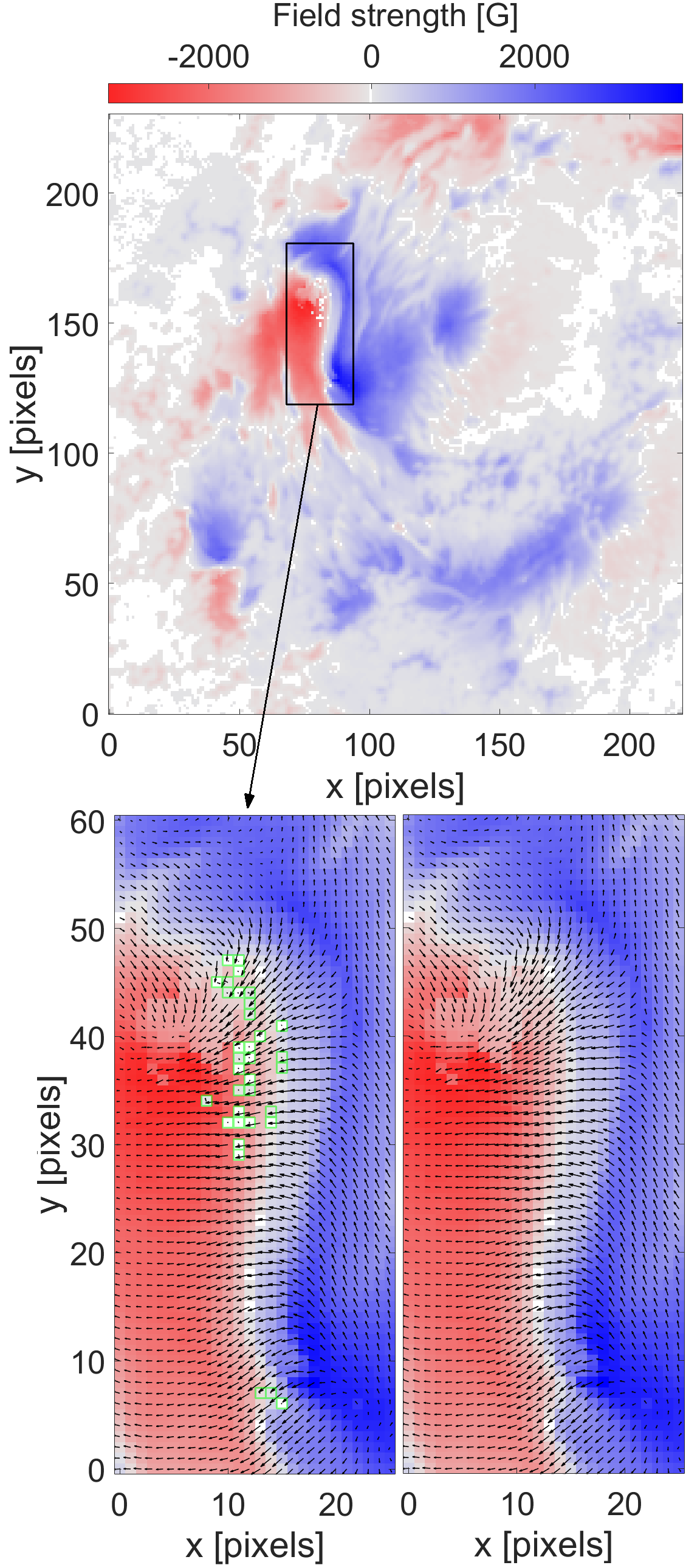}
\caption{
Magnetic field of the AR~12673 provided by {\it SDO}/HMI at 06 September 2017 03:34:42. Top row: Overview of the magnetic filed (pixel's color represents the value of the LOS-component). Bottom row, left: selected FOV with glitches (marked with green frames). Bottom row, right: the same FOV with removed glitches. Black arrows show direction and relative strength of the transversal component of the magnetic field. \label{fig:DeGlitch}}
\end{figure}

\subsection{SDO/HMI Vector Magnetograms: Removing Inversion Artifacts.}

Standard SDO/HMI vector magnetograms obtained on 06-Sep-2017 contain apparent artifacts in several pixels, which are most likely produced by an extremely fast emergence of the magnetic flux \citep{2017RNAAS...1a..24S}. 
These artifacts (glitches) are obvious from the bottom left panel of  Figure~\ref{fig:DeGlitch} (suspicious pixels are marked with green frames): both LOS and transversal components are totally different from most of their neighborhoods. This is more evident for the transversal component. We manually eliminated these glitches as follows: we identified a 'bad' pixel with the largest number of 'good' neighboring pixels and replaced the magnetic field components of this pixel with the average of it’s 'good' neighbors. Then we applied this algorithm recursively to all suspicious pixels. The result of such procedure is presented on the  Figure~\ref{fig:DeGlitch}, bottom right panel.

\subsection{Performing $\pi$-disambiguation.}
Vector magnetic field inverted from the Stokes profiles is intrinsically ambiguous.
The direction  of the transversal component can be measured only up to 180 degrees accuracy, meaning that
for every pixel in the magnetogram there are two possible directions of the transversal magnetic field. Thus, one has to perform a $\pi$-disambiguation procedure before using the magnetogram as a photospheric boundary condition for NLFFF reconstruction.
In this work, we employ two disambiguation methods: SFQ \citep{2014SoPh..289.1499R} and the Minimum Energy method \citep{1994SoPh..155..235M}.
Both methods give almost identical results in the strong field areas, which validates the $\pi$-disambiguations performed.

In the case of the magnetograms taken at 18:36:50~UT, there is an evident disambiguation artifact present in the outputs of both disambiguation methods as well as in the standard SDO/HMI data product.
To remove this artifact, we manually fix the field direction in the corrupted area and run the $\pi$-disambiguation procedures under this constraint.  This successfully removes the artifact.

\subsection{Maps of the GR source at 34\,GHz.} \label{S_radio_maps}
Standard data products of NoRH include daily images at 17 and 34\,GHz and 10-min images for 17\,GHz. To produce microwave images at 34\,GHz with a higher cadence (10 minutes) than available from the NoRH pipeline, % at specific time near Hinode SOT/SP magnetograms (03:34:42 UT), 
we use the software developed by the NoRH team and available at \url{https://solar.nro.nao.ac.jp/norh/archive.html}.
The image reconstruction has been performed using Hanaoka synthesis program that implements classical CLEAN algorithm \citep{1974A&AS...15..417H}.

These images consistently show high brightness temperature indicative of the gyroresonant emission process at least up to 34\,GHz. The centroid location of the 34\,GHz images, however, noticeably fluctuates from image to image within $\sim5\arcsec$ in a random direction. In Figure~\ref{fig:norh_obs} we shifted the 34\,GHz image by 3.5$\arcsec$ to the South, which is within the determined NoRH positioning accuracy.

\subsection{NLFFF Reconstruction of Coronal Magnetic field.} The photospheric vector magnetograms, with fixed artifacts and carefully resolved azimuth ambiguity,
were prepared for two time frames of interest, specifically, 03:34:42~UT and 18:36:50 UT on 6 September 2017. These magnetograms were employed as the bottom boundary conditions to perform NLFFF reconstruction of the coronal magnetic field.  Specifically, we employed  two different versions of the optimization method \citep{2000ApJ...540.1150W, 2004SoPh..219...87W} developed and validated by our team \citep{2017ApJ...839...30F}  using a full-fledged 3D MHD model as a proxy of a real AR. \citet{2017ApJ...839...30F} found that existing methods of the photospheric magnetic boundary condition preprocessing, proposed to remove the forced component of the magnetic field, in fact result in a corrupted height scale, \textbf{and} do not improve quality of the reconstruction. Therefore, we did not apply any preprocessing, but used the bottom boundary condition as is. Having unusually strong photospheric magnetic field implies a proportionally lower plasma beta, which further justifies the use of the force-free approximation. In what follows, we will validate our NLFFF reconstructions by comparison of the modeled values with those derived from the radio observations at the base of corona, where the force-free approximation holds much better. Both reconstructions resulted in very similar coronal magnetic structures, which provides confidence in the obtained solutions. To avoid any confusion, we use here only one of the solutions; namely, the one obtained with the weighted optimization  code \citep{2017ApJ...839...30F}.

\subsection{Chromospheric and coronal thermal model.} 
Presence of a strong magnetic field of $\sim$4\,kG at the coronal heights, where the thermal plasma is hot, is an absolute prerequisite of having a bright gyroresonant radio source at 34\,GHz as the one reported in Section~\ref{S_radio_maps}. The magnetic model alone is, however, insufficient to compute the radio brightness needed for a meaningful model-to-data comparison, which in addition requires a thermal model on top of the magnetic one. Thus,to compute simulated radio emission from the model,
\textbf{its} volume has to be filled in with a radiating thermal plasma.
Within the GX Simulator methodology \citep{2018ApJ...853...66N}, this process is called the \textit{volume reprocessing}. During the reprocessing, the length and the mean value of the magnetic field along the field line associated with a given volume element (voxel)  are computed for each voxel of the 3D magnetic model.  
Then, a parametric model of the coronal plasma heating is applied to every voxel crossed by a closed line of the magnetic field obtained from the reconstruction. A thermal model of the chromosphere, which has a nonuniform adaptive height spacing needed to resolve the transition region, is added at the bottom part of the data cube following a photospheric mask obtained from a joint analysis of the white light and LOS magnetic field maps \citep{2018ApJ...853...66N}. 
This way, the magnetic \lq\lq skeleton\rq\rq\ is getting filled with a plasma with appropriate temperature and density forming a thermal structure of our model.

\section{Discussion}

Starting from the photospheric vector magnetograms, we built 3D coronal magnetic models using nonlinear force-free field (NLFFF) reconstruction, populated these data cubes with coronal and chromospheric thermal plasma \citep{2018ApJ...853...66N}, and validated them by comparison of the simulated microwave emission with  the microwave imaging data  
at 17 and 34\,GHz. Almost perfect match between the simulated and observed images (see Figure~\ref{fig:norh_obs}) validates both magnetic and thermal structure of our 3D model.

\begin{figure*}
	\begin{center}
		\includegraphics[width=0.5\linewidth]{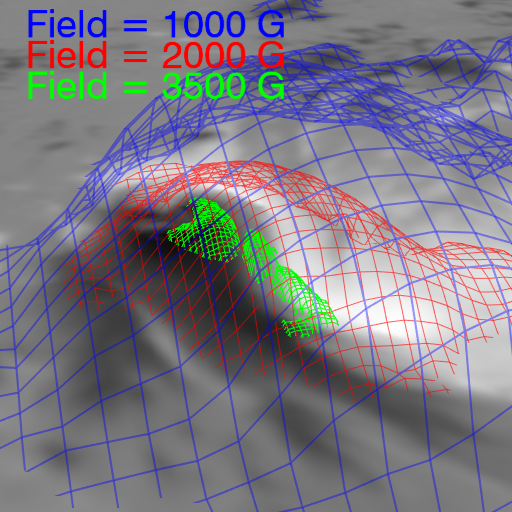}\includegraphics[width=0.5\linewidth]{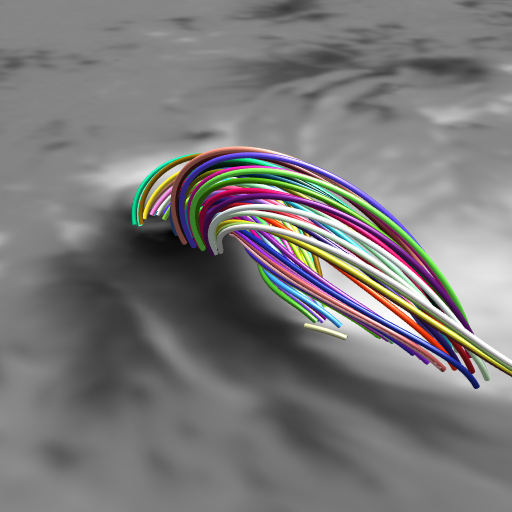}
	\end{center}
		\caption{
		Iso-gauss layers corresponding to the magnetic field strength of 1000\,G (blue), 2000\,G (red), and 3500\,G (green). The right panel shows the magnetic field lines rooting at the photospheric areas where the magnetic field exceeds 3500\,G\label{fig:hats}. See also 3D animations in supplemented materials.}
\end{figure*}

Figure~\ref{fig:hats}a shows a 3D representation of the coronal magnetic field by displaying iso-gauss surfaces of 3,500\,G 
(green wired surface), 2,000\,G (red wired surface), and 1,000\,G (blue wired surface). This figure demonstrates that the kilogauss magnetic field occupies a highly significant coronal volume at the active region, extending to much higher heights than in a typical case. Figure~\ref{fig:hats}b gives a complementary view of the magnetic structure by showing a flux tube of the strongest, highly twisted magnetic field forming a flux rope in the core of the active region.

\begin{figure}
	\begin{center}
        \includegraphics[width=1.\linewidth]{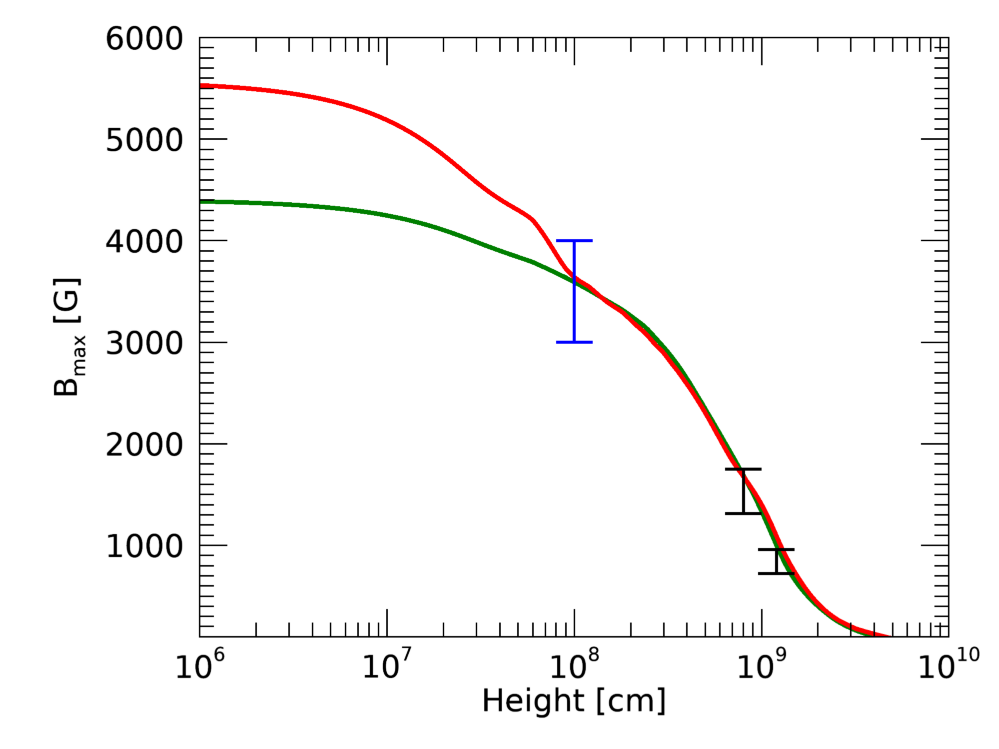}
	\end{center}
		\caption{
		Height dependence of the strongest magnetic field obtained from the NLFFF models for 03:36 UT (green line) and 18:36 UT (red line).
		Magnetic field estimate at the base of the corona obtained from NoRH data is shown with blue vertical dash. The strongest coronal magnetic field ever reported \citep{2006ApJ...641L..69B} based on VLA observations above the limb are shown with black vertical dashes.
		\label{fig:Bmax_vs_height}}
\end{figure}

To quantify the height dependence of the coronal magnetic field, Figure~\ref{fig:Bmax_vs_height} displays the dependence of the maximum magnetic field value at a given height (the green line) along with  the  estimate for the magnetic field  at the base of the corona derived from the GR emission at 34\,GHz. We also plot the historically largest measurements of the magnetic field obtained by \citet{2006ApJ...641L..69B} at two different heights from VLA observations above the limb. It is interesting that the upper bounds of these measurements  match our model surprisingly well. This indicates that AR 12673 did show the strongest, record-braking coronal magnetic field. However, given the proximity of the VLA data points to our curve that displays the largest magnetic field vs height, the comparably strong coronal magnetic fields might be more common than has been appreciated so far.  To check this expectation, we looked at the historical daily record of the NoRH data at 34\,GHz and indeed found many cases showing unexpectedly bright radio sources indicative of extremely strong coronal field. Some of these cases correspond to the strongest sunspot magnetic field reported by \citet{2018ApJ...852L..16O}. A detailed analysis of this data set is underway and will be published elsewhere shortly.

\citet{2018RNAAS...2a...8W} reported strong photospheric magnetic field of 5570~G at AR 12673 based on Goode Solar Telescope (GST) data, detected roughly half day after our observations of the bright gyroresonant source. To investigate whether this enhancement of the photospheric magnetic field had any effect on the coronal magnetic field, we obtained the photospheric vector magnetic field using Hinode SOT/SP  data from a restricted field of view at 18:36:50~UT that demonstrated a consistently strong magnetic field up to 5,700~G at the light bridge, embedded this magnetogram into a larger field-of-view SDO/HMI magnetogram, and produced a NLFFF reconstruction of the coronal magnetic field. Then, we computed the largest magnetic field at each height and plotted it in Figure~\ref{fig:Bmax_vs_height} in red. Interestingly, the red curve exceeds the level of the green one (obtained for 03:34:42~UT) only over a very restricted range of the heights low in the solar atmosphere, while then merges the green one. 

This finding suggests that the strong transverse photospheric magnetic field, reported by
\citet{2018RNAAS...2a...8W}, closes at the photosphere or low chromosphere, but does not propagate upward to the corona. In contrast, a bigger area of a slightly weaker, but still very strong magnetic field of about 5,000~G, that persists in this active region during the day of Sep-06 or even longer has a significant imprint in the coronal magnetic field.

\section{Conclusions}

\begin{figure*}
	\begin{center}

\includegraphics[width=1.\linewidth]{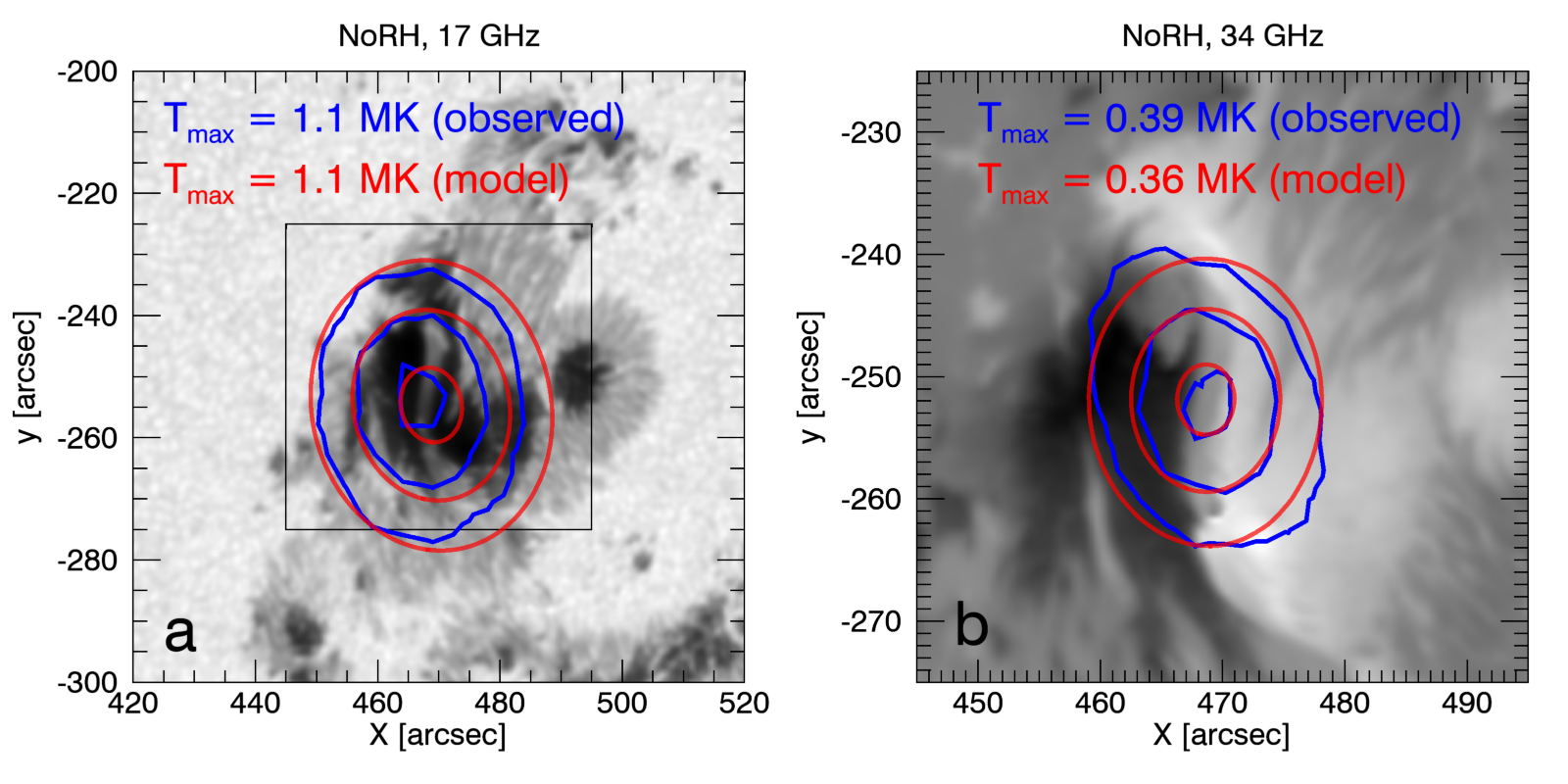}
	\end{center}
		\caption{
		Continuum white light emission (a) and LOS magnetogram (b) obtained by SDO/HMI on 2017-06-09 03:34:42. Blue contours show microwave radio emission observed with NoRH on 03:30:01~UT at 17\,GHz (a) and 34\,GHz  (b).
		The synthetic radio maps calculated from our 3D magneto-thermal %NLFFF 
		model are shown by the red contours.
		In all cases, the contours are plotted at the levels of  20, 50 and 90\% of the peak brightness temperature.
		The peak brightness temperatures are given by the labels in the top left corner of each panel.
		The black rectangle shown on the left panel indicates the field of view of the panel (b).
		Note that the radio sources project on the light bridge (a bright vertical  feature seen in panel (a) at the center of the radio source), which is elongated along the polarity inversion line seen in the magnetic field LOS map shown in panel (b).
	}\label{fig:norh_obs}
\end{figure*}

It is extremely exciting that AR~12673 shows a unique, long-living bright GR source at 34~GHz (see  Figure~\ref{fig:norh_obs}, right panel), which is twice larger than the largest frequency at which a GR source has ever been reported indicative of an unexpectedly strong coronal magnetic field in this case. In addition, we found that the historically strongest coronal magnetic field values, obtained by \citet{2006ApJ...641L..69B} at two different heights from VLA observations above the limb, match the obtained here height dependence of the strongest magnetic field remarkably well.  This implies that strong coronal magnetic fields, comparable to that reported here, might be a far more common phenomenon than has been appreciated so far.

\acknowledgments
This work was supported by the by the RFBR research grant 18-32-20165 \verb"mol_a_ved" (S.A., A.S., I.M.) and the by program of Basic Research No. II.16 (S.A., I.M.).
G. F. was supported in part by NSF grant AST-1820613
and NASA grants 80NSSC18K0667 and 80NSSC19K0068 to the
New Jersey Institute of Technology.
We also thank NASA SDO/HMI and NoRH teams.

\bibliographystyle{apj}

\end{document}